\newif\ifmulticol	\multicoltrue
\newif\ifshowgit	\showgittrue		% switches footer on/off
\newif\ifgitlocal	\gitlocalfalse		% use local file gitHeadLocal.gin
\newif\ifbiblatex	\biblatexfalse		% defaults to bibtex if false
\newif\ifbibnum		\bibnumtrue 		% num => superscripts, otherwise auth date
\newif\iflineno		\linenofalse
\newif\iftoc		\tocfalse

\newif\iflucida		\lucidafalse
\newif\ifcm			\cmfalse
\newif\iflibertine	\libertinefalse
\newif\ifcharter	\chartertrue

\newcommand*{\secnumdepth}{0}			% sets secnumdepth in \mymaketitle
\newcommand*{\tocdepth}{2}				% sets tocdepth in \mymaketitle
\newcommand*{\reftitle}{References}		% title for references section
\newcommand*{\mytitleskip}{-0.2in}		% change skip after title

\multicoltrue\showgittrue\gitlocaltrue\biblatexfalse\bibnumtrue

\newcommand*{\mydocfontsize}{\ifcharter11pt\else\iflibertine11pt\else10pt\fi\fi}
\newcommand*{\setcol}{\ifmulticol twocolumn\else onecolumn\fi}

\documentclass[\mydocfontsize,\setcol]{article} 

\newcommand*{\pdftitle}{Metabolic heat in microbial conflict and cooperation}

\usepackage{siunitx}

\input heat.sty
\newcommand*{\dC}{\si{\degreeCelsius}}

\usepackage{bm}

\DeclarePairedDelimiter\abs{\lvert}{\rvert}
\DeclarePairedDelimiter\norm{\lVert}{\rVert}
\DeclarePairedDelimiter\angb{\langle}{\rangle}
\DeclarePairedDelimiter\lrb{\lbrack}{\rbrack}
\DeclarePairedDelimiter\lr{\lparen}{\rparen}
\DeclarePairedDelimiter\lrbr{\lbrace}{\rbrace}

\makeatletter
\let\oldabs\abs \def\abs{\@ifstar{\oldabs}{\oldabs*}}
\let\oldnorm\norm \def\norm{\@ifstar{\oldnorm}{\oldnorm*}}
\let\oldangb\angb \def\angb{\@ifstar{\oldangb}{\oldangb*}}
\let\oldlrb\lrb \def\lrb{\@ifstar{\oldlrb}{\oldlrb*}}
\let\oldlr\lr \def\lr{\@ifstar{\oldlr}{\oldlr*}}
\let\oldlrbr\lrbr \def\lrbr{\@ifstar{\oldlrbr}{\oldlrbr*}}
\makeatother

\newcommand*{\boldrule}{\hrule height 1.2pt}
\newcommand*{\noterule}{\medskip\boldrule\medskip}	% for notes

\newcommand*{\mytitle}{\pdftitle}

\newcommand*{\myauthor}{Steven A.\ Frank}
\newcommand*{\myaddress}{Department of Ecology and Evolutionary Biology, University of California, Irvine, CA 92697--2525, USA}
\newcommand*{\mycontact}{\href{https://stevefrank.org}{https://stevefrank.org}}

\newcommand*{\mykeywords}{Thermoregulation, microbial metabolism, overflow metabolism, biofilms, public goods, social evolution, ecological competition, fever, bacteriophage defense}
\setabstract{-0.07}{\iftoc-2.0\else1.22\fi}{%
Many microbes live in habitats below their optimum temperature. Retention of metabolic heat by aggregation or insulation would boost growth. Generation of excess metabolic heat may also provide benefit. A cell that makes excess metabolic heat pays the cost of production, whereas the benefit may be shared by neighbors within a zone of local heat capture. Metabolic heat as a shareable public good raises interesting questions about conflict and cooperation of heat production and capture. Metabolic heat may also be deployed as a weapon. Species with greater thermotolerance gain by raising local temperature to outcompete less thermotolerant taxa. Metabolic heat may provide defense against bacteriophage attack, by analogy with fever in vertebrates. This article outlines the theory of metabolic heat in microbial conflict and cooperation, presenting several predictions for future study.
}

\begin{document}

\mymaketitle

%% 1st parm is skip on left column at start of TOC, 2nd param is skip after TOC
\iftoc\mytoc{-24pt}{\newpage}\fi

\section{Introduction}

Metabolic heat may play an important role in microbial conflict and cooperation. On the conflict side, microbes often differ in their temperature optima \autocite{alster18a-meta-analysis}. A microbe that raises the local temperature closer to its own optimum gains a growth advantage over relatively thermophobic competitors \autocite{goddard08quantifying}. 

On the cooperative side, aggregates may retain metabolic heat and gain a growth rate advantage \autocite{tabata13measurement}. Internal cells in an aggregate potentially benefit by generating excess heat, the energy cost reducing their own growth but stimulating faster growth among neighboring genetic relatives. Cool environments with slow heat dissipation favor cooperative thermogenesis.

The theory builds on three assumptions. First, temperature influences fitness. Second, individual and group traits can modulate heat production and heat flow. Third, heat flow affects local temperature and thus the fitness of neighbors.

I discuss each assumption. I then turn to predictions.

\section{Fitness consequences: competition}

Competition requires that taxa differ in their temperature response. Types with relatively higher temperature optima or greater tolerance to heat can potentially gain an advantage by warming the local environment. 

Alster et al. \autocite{alster18a-meta-analysis} quantified temperature response in terms of thermodynamic variables that determine reaction rates. Thermodynamic measures of reaction rates are not direct measures of fitness. However, differences in the temperature sensitivity of metabolic processes likely influence temperature differences in growth rate and metabolic efficiency.

Alster et al. \autocite{alster18a-meta-analysis} focused on three variables. The optimum temperature maximizes reaction rate. The heat capacity determines the breadth of the temperature response curve, with greater heat capacity corresponding to a broader temperature response curve and less overall sensitivity to temperature variability. Maximum temperature sensitivity defines the point at which the reaction rate changes most rapidly with respect to temperature. 

Literature meta-analysis yielded 353 response curves across diverse microbial groups \autocite{alster18a-meta-analysis}. The distribution of optimum temperature is approximately a right-skewed Gaussian shape with a mean and standard deviation of $29.4\pm10.1$\dC. Heat capacity and maximum temperature sensitivity also vary widely.

Many laboratory studies have measured fitness at different temperatures \autocite{bennett07an-experimental,chen10thermal,caspeta15thermotolerant,yung15thermally}. Growing microbes at higher or lower temperatures often causes an evolutionary shift in temperature response, demonstrating lability of thermotolerance. Protein thermosensitivity likely explains a significant part of the variability in the temperature response curves of taxa \autocite{ghosh10cellular,chen17thermosensitivity}.

Overall, variation in observed temperature response suggests wide scope for using metabolic heat as a competitive weapon.

\section{Fitness consequences: cooperation}

The potential for cooperation requires that microbes sometimes live in habitats below their optimum temperature. When below the optimum, excess heat can be a shareable public good that is costly to produce and potentially beneficial to neighbors. I found only one study of metabolic heat used to raise local temperature for colony benefit \autocite{tabata13measurement}. 

Several examples suggest that increased local temperature could be advantageous in cold environments.

Some organisms use dark pigmentation to raise cellular temperature. Cordero et al. \autocite{cordero18impact} showed that pigmentation increases in yeast with latitude. That increase suggests that high latitude taxa gain from raising their temperature above the ambient level. When grown in the lab at 4\dC\ under light, melanized \textit{Crytococcus neoformans} gained a growth advantage relative to nonmelanized variants, but at 23\dC\ the melanized form suffered increased thermal stress.

Most of the biosphere is permanently cold, including alpine, arctic, and oceanic habitats \autocite{rodrigues08coping}. Cold-adapted microbes  \autocite{damico06psychrophilic} occupy these habitats down to about $-20\dC$. Estimates suggest counts of approximately $10^5$ and $10^6$ cells \si{\per\ml} in Arctic ice pack and Antarctic sea ice, respectively \autocite{brinkmeyer03diversity}. Smaller counts have been observed in deep ice cores \autocite{price04temperature}.

Among taxa that could be cultured \autocite{rodrigues08coping}, most isolates from cold habitats survived or grew at cold temperatures but reproduced most quickly at 20--25\dC. Only a few isolates grew fastest at cool temperatures of 10--15\dC. Thus, the capture and sharing of local metabolic heat may be particularly valuable in cold habitats.

\begin{figure*}[t]
\centering
\vskip0.05in
\includegraphics[width=0.6\hsize]{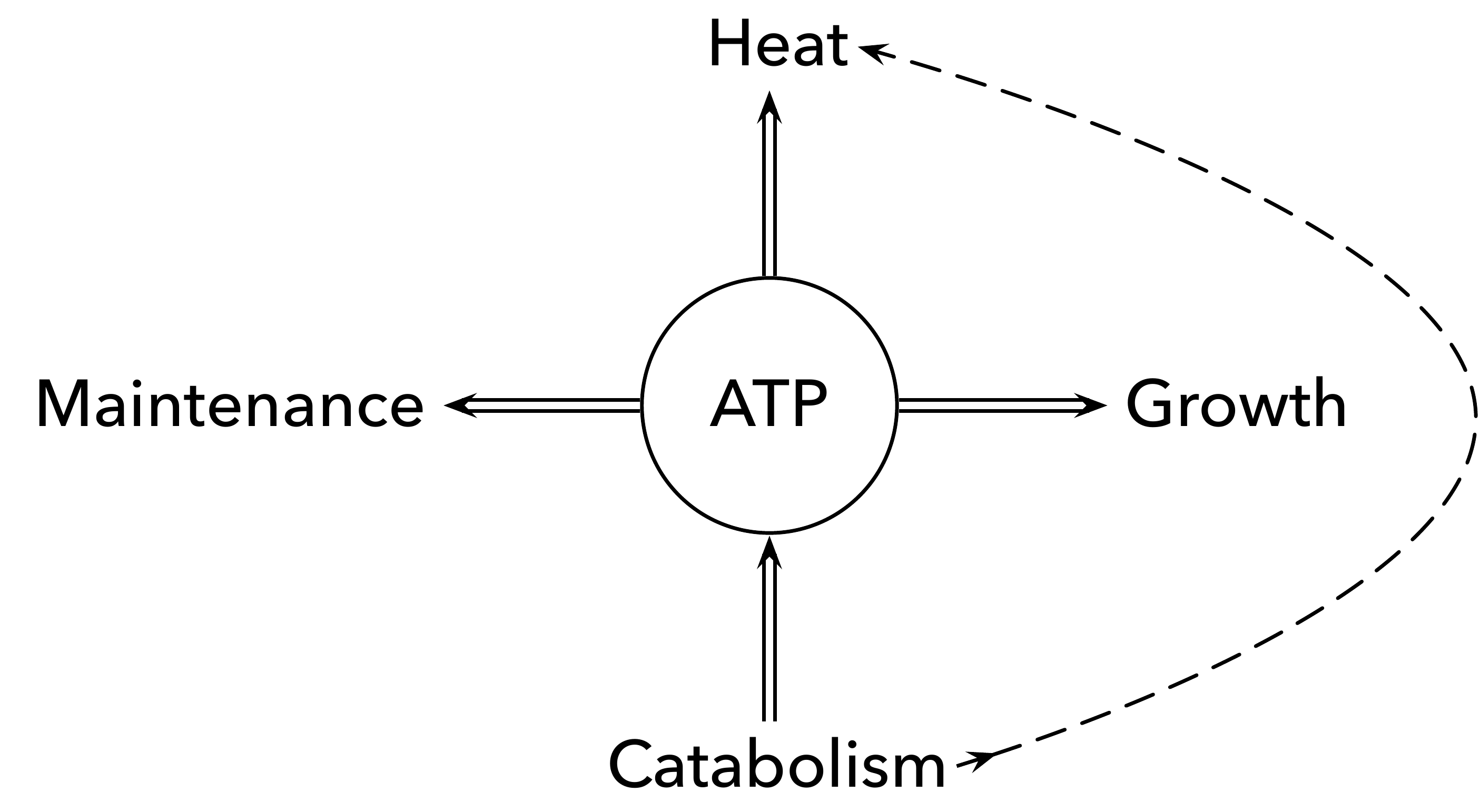}
\vskip0.1in
\caption{Intrinsic versus excess metabolic heat production. Catabolism of food yields ATP and intrinsically generated metabolic heat (dashed curve). A cell may use the free energy added to the ATP/ADP disequilibrium to drive cellular maintenance and growth. Alternatively, the ATP/ADP disequilibrium can drive futile biochemical processes that may function to relieve biochemical imbalances or to generate excess beneficial heat. Any benefit of excess heat trades off against lost ATP/ADP disequilibrium to drive maintenance and growth. Alternatively, cells can generate excess metabolic heat by reducing the ATP produced during catabolism and increasing the direct heat generated along the dashed curve. For example, allowing proton motive force to cross the membrane without driving ATP synthase increases the heat production associated with the ion flow, as can happen in the electron transport chain or similar processes. Approximate calculations intended only to provide a rough sense of magnitude suggest that a cell aerobically respiring its own weight in glucose produces enough intrinsic metabolic heat to raise cellular temperature by about 2\dC\ and enough ATP that could be used to generate additional excess metabolic heat to raise temperature by another 2\dC. Cells that bypass ATP production and act almost purely as excess heat producers may be able to increase maximum metabolic rate because they have a larger free energy gradient between glucose and the final metabolic products. Modified from Figure 1 of Russell \autocite{russell07the-energy}.} 
\label{fig:futileCycle}
\end{figure*}

\section{Individual and group traits}

Individuals may contribute heat by excess thermogenesis. Groups may retain heat by aggregation and by insulation.

Cellular aggregation is perhaps the simplest trait. I did not find studies of microbes that consider individual and group traits in terms of conflicting and cooperative aspects of thermoregulation. The closest analogy to my argument comes from huddling behavior in birds and mammals to retain heat. 

Haig \autocite{haig08huddling:,haig10the-huddlers} noted that heat is a public good in a vertebrate huddle. Heat generators pay the cost of production. The benefit is shared by all neighbors. Individuals can exploit warm neighbors by reducing their own heating budget. In broods, siblings and parents have various conflicting and cooperative interests with regard to heat.

Familial conflicts over physiological traits often associate with genomic imprinting in mammals \autocite{haig10transfers}. Several imprinted genes in mice and humans influence thermogenesis and follow the common pattern for familial conflict \autocite{haig08huddling:,crespi20why-and-how-imprinted}. 

In microbes, aggregation by intercellular adhesion occurs widely. Cells may also aggregate by surface attachment and by active movement toward groups. Many possible costs and benefits of cellular aggregation occur \autocite{grosberg07the-evolution,koschwanez11sucrose,smith14fruiting,ratcliff15origins,kuzdzal-fick19disadvantages}. However, heat in microbial aggregates has not been widely discussed.

In addition to aggregation, groups may also retain heat by secreting extracellular insulation. The idea that extracellular secretions function as insulation for microbial thermoregulation has not been widely discussed.

Biofilms combine aggregation and insulation. Apart from the one study mentioned above  noting that aggregates may beneficially raise their temperature \autocite{tabata13measurement}, I did not find discussion of increased temperature as an adaptive benefit of biofilms. The idea has likely been mentioned, but is not widely considered.

Heat may be used as a weapon against relatively thermophobic competitors or invaders. Godard's \autocite{goddard08quantifying} suggestion that \textit{Saccharomyces cerevisiae} may use metabolic heat as a competitive weapon against other species is the only clear statement that I found. I did not find any literature on microbes that use metabolic heat as a defense against invading bacteriophage. 

With regard to defense, fever in vertebrates \autocite{evans15fever} and social insects \autocite{starks00fever} provides an analogy. Those organisms sometimes raise their temperature to control microbial invaders. Observations suggest that Japanese honey bees surround invading Asian giant hornets (\textit{Vespa mandarinia}) and generate excess heat and CO$_2$ to kill their relatively thermophobic enemy \autocite{sugahara09heat}.

These various benefits of excess heat require that some individuals in the group spend metabolic energy on heat production (Fig.~1). Microbes have metabolic flux pathways that seem designed to dissipate excess ATP without driving any anabolic processes. Those energy spilling reactions release significant heat, sometimes associated with a futile cycle of proton flux through the cell membrane \autocite{russell07the-energy}. 

A few possible functions for futile cycles have been mentioned, such as correcting thermodynamic imbalance \autocite{von-stockar99does}. However, the Tabata et al.\ article \autocite{tabata13measurement} is the only one I found that suggests heat generation may itself be a benefit.

If free energy limits reproduction, then individuals that generate excess heat may be reducing their own reproduction in favor of the group-level benefit shared by neighbors. With heat as a public good \autocite{west07the-social}, competitive nonproducers could gain a growth advantage against cooperative heat producers.

In dense, energy-rich environments that dissipate heat relatively slowly, metabolic heat can raise local temperatures beyond the optimum for growth. Excreting catabolic intermediates such as lactate, acetate or ethanol may reduce heat production to keep temperatures below stressful levels. Protection against overheating provides an alternative explanation for the puzzle of overflow metabolism \autocite{warburg56on-the-origin,postma89enzymic,wolfe05the-acetate}.

\section{Heat flow and spatial scale}

Metabolic heat can alter local temperature and potentially be important in conflict and cooperation. However, local heat must dissipate sufficiently slowly to play an important role. Here, ``slowly'' means the scaling of heat dissipation relative to the rate of other processes.

For example, does excess heat dissipate slowly enough that it can raise the rates of metabolic reactions and the growth rate of neighbors? Can excess heat be sufficiently concentrated to be used as a weapon that reduces the growth rate of relatively thermophobic competitors? What aspects of cellular aggregation and biofilm properties retain heat sufficiently to raise growth rate? How do changes in heat flow trade off against changes in the flow of other resources? How do larger-scale biophysical aspects of a habitat interact with smaller-scale intercellular processes to affect overall heat conductance?

Habitats vary in thermal properties. For example, water content and particle size significantly influence heat flow in soils \autocite{usowicz13effects}. Water absorbs and dissipates heat more rapidly than does air. Convective flow may often dominate in the movement of heat. Still habitats may therefore be better candidates for local concentration of heat.

Many articles consider whether temperature gradients can be maintained within single cells \autocite{inada19temperature,suzuki20the-challenge}. The current consensus suggests that significant intracellular gradients are unlikely because heat diffuses too quickly within cells and across membranes \autocite{balaban20how-hot-are-single,oyama20single-cell}.

My arguments concern multicellular interactions and insulated environments, so the single-cell controversy is not directly relevant. But the technical issues of measurement \autocite{braissant10use-of-isothermal} may be important for studies of heat flow within multicellular aggregations. Future improvements in technology will likely enhance spatial resolution, which may improve the tracking of heat flow over the spatial scales at which conflict and cooperation play out.

With regard to the single-cell scale and cooperation, Dunn \autocite{dunn17some} suggested that the first step toward eukaryotes arose when an archaeal cell acquired a bacterial symbiont as a source of internal cellular heat. The additional intracellular heat might have allowed thermophilic archaea to migrate to colder environments by a form of endothermy. However, the most recent single-cell studies mentioned above tend toward rejecting significant temperature differences between organelles, the cytosol, and the extracellular environment. 

\section{Discussion}

The potential role of metabolic heat in microbial conflict and cooperation follows from basic observations and simple ideas. However, no evidence directly supports the theory, perhaps because the problem has rarely been discussed or studied. Several broad predictions summarize key points and future applications.

Relatively cold habitats more strongly favor excess metabolic heat to raise local temperature.

Habitats that dissipate heat more slowly favor the benefits of local heat production more strongly.  

Cellular aggregation and extracellular insulation to retain local heat are more strongly favored as the growth rate benefits from heat become more valuable competitively.

The more genetically distinct cells are in an aggregation, the more likely that some cells do not contribute to costly heat production (the public goods dilemma). 

In cellular aggregations, internal cells are more likely to generate excess heat because their heat production is protected by greater insulation than peripheral cells.

In biofilms, some of the observed spatial variation in gene expression between cells \autocite{lenz08localized,besharova16diversification} may associate with internal heat production and external insulation. 

Species with broader thermal tolerance and higher optimum temperature are more likely to use local heat as a competitive weapon.

Competitive heat generation favors competitors to raise their thermotolerance, which alters the selective pressure on competitive heat production, leading to game-like dynamics.

Species with broader thermal tolerance are more likely to use excess heat generation as a defensive fever response against invading bacteriophage.

The benefit of competitive and defensive heat generation may often increase with the number of cells that cooperate to create a thermal weapon, suggesting a link to quorum sensing.

The benefit of competitive and defensive heat generation rises with the tendency of the aggressors to surround their foe.

In dense environments that tend to overheat, microbes may secrete catabolic intermediates in overflow metabolism to reduce heat generation.

In summary, heat plays a primary role in the rate processes of life. Various individual and group traits of heat generation, cellular aggregation, and extracellular insulation may influence aspects of conflict and cooperation in microbial communities.

\section*{Acknowledgments}

\noindent The Donald Bren Foundation, NSF grant DEB-1939423, and DoD grant W911NF2010227 support my research.

%\vfill\eject

\mybiblio	% uses main.bib by default, add other bibs as needed

\addcontentsline{toc}{section}{Appendix}

% used cuted package strip env to force balancing of columns
%\ifmulticol\begin{strip}\hbox{\null}\end{strip}\hbox{\null}\fi

\end{document}